# Spin Transfer Torque as a Non-Conservative Pseudo-Field

Sayeef Salahuddin*, Deepanjan Datta and Supriyo Datta

*School of Electrical and Computer Engineering and NSF Center for Computational*

*Nanotechnology (NCN), Purdue University, West Lafayette, IN 47906*

*Present address: Electrical Engineering and Computer Science, UC Berkeley, CA-94720

**Abstract:**

In this paper we show that the spin transfer torque can be described by a pseudo magnetic field, proportional to the magnetic moment of the itinerant electrons that enters the Landau-Lifshitz-Gilbert equation in the same way as other external or internal magnetic fields. However, unlike an ordinary magnetic field, which is always conservative in nature, the spin torque induced 'pseudo field' may have both conservative and non-conservative components. We further show that the magnetic moment of itinerant electrons develops an out-of-plane component only at non-equilibrium and this component is responsible for the 'Slonczewski' type switching that acts against the damping and is always non-conservative. On the other hand, the in-plane components of the pseudo field exist both at equilibrium and out-of-equilibrium, and are responsible for the 'field like' term. For tunnel based devices, this term results in lower switching current for anti-parallel (AP) to parallel (P) switching compared to P to AP, even when the torque magnitudes are completely symmetric with voltage.

## 1. Introduction

Spin torque devices [1, 2] that switch the magnetization of small magnets with spin polarized currents without any external magnetic field, have stirred tremendous interest due to their potential application as non volatile memory and also as nanoscale microwave oscillators. Although the concept of spin transfer torque has been demonstrated by a number of experiments [3, 4], quantitative measurement of spin transfer torque has been achieved only very recently [5, 6, 7]. All these measurements show a significant 'field-like' or out-of-plane torque in addition to the original in-plane torque predicted by Slonczewski [1]. This is very different from metallic channel based devices where the field like term is minimal. Recent theoretical studies have also shown the field like term to be significant in tunnel based devices [8, 9, 10, 11]. However, the details of how this field-like torque can affect the switching behavior is yet to be understood properly [5, 6, 7, 12, 13].

In this paper we first show that spin torque can be described by a pseudo magnetic field proportional to the net magnetic moment $\vec{\mu}$ of the itinerant electrons, (normalized to the Bohr magneton $\mu_B$) providing a natural relationship between Slonczewski and field like terms:

$$\frac{d\hat{m}}{dt} = -|\gamma|\,\hat{m}\times\left(\vec{H}_{eff} + k\,\vec{\mu} - \frac{\alpha}{|\gamma|}\frac{d\hat{m}}{dt}\right) \tag{1}$$

Eqn. (1) is the central result of this paper and is derived in Section 2, starting from the Gilbert form of the LLG equation and introducing the spin-torque in terms of $\vec{\mu}$ obtained from non-equilibrium Green function (NEGF) formalism for the conduction electrons. Note that $k > 0$, so that the pseudo field $\left(k\,\vec{\mu}\right)$ is in the same direction as $\vec{\mu}$ and enters Eqn. (1) just like other magnetic fields included in $\vec{H}_{eff}$. This may seem surprising; since it is well-known that spin-torque leads to phenomena like coherent precession that do not arise from ordinary magnetic fields. We show in section 3 that such phenomena can also be understood in terms of Eqn. (1) once we note that the pseudo-field $k\vec{\mu}$ representing the spin-torque has both a conservative component like the conventional magnetic fields included in $\vec{H}_{eff}$ and also a non-conservative component that makes the curl of overall $\vec{\mu}$ to be non-zero: $\vec{\nabla}_m \times \vec{\mu} \neq 0$; (Note that $\vec{\nabla}_m \times \vec{H}_{eff} = 0$ ).

We show that the out-of-plane component of $(k\vec{\mu})$ is responsible for the Slonczewski term and is always non-conservative. On the other hand, the in plane components give the field like term and can introduce asymmetry in switching currents for opposite polarity in the voltage bias. Specifically, we shall show that for tunnel based devices, this field like torque can result in a lower switching voltage for AP to P switching compared to P to AP, even when the torque magnitudes are completely symmetric with voltage. This can be understood by noting that for tunneling devices, the in plane component of the pseudo field (responsible for the field like term) remains conservative even away from equilibrium, and thus acting like an ordinary magnetic field parallel to the direction of the fixed magnet that helps switching from AP to P while hindering P to AP transition.

**2. Spin-Torque as a Pseudo Field**

A typical spin torque device is shown schematically in Fig. 1. Left contact is the fixed ferromagnet having magnetization along $\vec{M}$. Right contact is soft layer and its magnetization points along $\vec{m}$ which is free to rotate in easy (z-x) plane. An insulating layer separates the ferromagnetic contacts. Following Gilbert's prescription, we write, the rate of change of the direction $\hat{m}$ of the magnetization as

$$\frac{d\hat{m}}{dt} = -|\gamma|\hat{m} \times \left(\vec{H}_{eff} - \frac{\alpha}{|\gamma|}\frac{d\hat{m}}{dt}\right) + \frac{\vec{T}}{M_s V} \quad (2)$$

where the spin torque $\vec{T}$ is obtained by integrating the divergence of the spin current $\vec{J}_s$ carried by the conduction electrons over the volume of the magnet. Below we will use the non-equilibrium Green's function formalism to show that

$$\vec{T} = -\mu_B \int dV \vec{\nabla}.\vec{J}_s = -(\mu_B \Delta / \hbar)\hat{m} \times \vec{\mu} \quad (3)$$

where $\vec{\mu}$ is the magnetic moment of the conduction electrons (normalized to $\mu_B$) and $\Delta$ is the energy splitting of the conduction electrons due to the exchange interaction with the localized spins that comprise the magnet. Combining Eqns (2) and (3) we obtain our central result stated earlier in Eqn. (1) with $k = \mu_B \Delta / |\gamma|\hbar M_S V = \Delta / 2M_S V$.

***Proof of Eqn. 3:*** We start from the expression for the (2x2) operator representing $\vec{\nabla}.\vec{J}$ at site '$j$' in a discrete representation (see Eqn. (8.6.3), page 317, [15]) for the conduction electrons $\vec{\nabla}.\vec{J} = [HG^n - G^n H]/i\hbar$. $G^n(\equiv -iG^<)$ is the (2x2) correlation matrix at site '$j$' and the Hamiltonian $[H]$ is given by $H = hI + (\vec{\sigma}.\hat{m})\Delta$, where $hI$ is the spin-independent part and $(\vec{\sigma}.\hat{m})\Delta$ is the spin-dependent part arising from the exchange interaction with the magnet pointing along $\hat{m}$, with $I$ being a 2x2 identity matrix and $\vec{\sigma}$ representing the Pauli spin matrices.

The divergence of the spin-current is obtained from the $\vec{\nabla}.\vec{J}$ operator

$$[\vec{\nabla}.\vec{J}_S] = \frac{1}{i\hbar}\int dE \left\{Trace\left[\sigma_i \left(HG^n - G^n H\right)\right]\right\} \quad (4)$$

and substituting for $H$, we get (note: $\varepsilon_{ijk}$ is the Levi-Civita antisymmetric tensor)

$$\left[\vec{\nabla}.\vec{J}_S\right]_i = \frac{\Delta}{i\hbar}\sum_j m_j \int dE \left\{Trace\left[\sigma_i\sigma_j G^n - \sigma_i G^n \sigma_j\right]\right\} = \frac{\Delta}{\hbar}\sum_j \int dE \left\{\varepsilon_{ijk} m_j Trace\left[\sigma_k G^n\right]\right\} \quad (5)$$

so that the spin-torque is given by

$$T_i = \mu_B \int dV \left[\vec{\nabla}.\vec{J}_S\right]_i = \frac{\mu_B \Delta}{\hbar}\int dV \sum_j \int dE \left\{\varepsilon_{ijk} m_j Trace\left[\sigma_k G^n\right]\right\} \quad (6)$$

Defining $\mu_k \equiv -\int dV \int dE\, Trace\left[\sigma_k G^n\right]$ as the magnetic moment (normalized to $\mu_B$) of the conduction electrons, we obtain $T_i = -(\mu_B \Delta/\hbar)\sum_j \varepsilon_{ijk} m_j \mu_k$ which is the same as $\vec{T} = -(\mu_B \Delta/\hbar)\hat{m}\times\vec{\mu}$ as stated above in Eqn. (3). This completes our proof of Eqn. (1). Note that this expression for torque is consistent with previous studies [9,10,11].

## 3. Relation to the standard form

It is shown in the *Appendix A* that if the conduction electrons are in equilibrium then the spin density $\vec{\mu}$ can be written in the form,

$$\vec{\mu}(eq) = a'\hat{m} + b'\hat{M} \quad (7)$$

but away from equilibrium, the spin density remarkably develops an additional out-of plane component that is perpendicular to the magnetization of both magnets:

$$\vec{\mu} = a\hat{m} + b\hat{M} + c\left(\hat{m}\times\hat{M}\right); \quad (8)$$

so that from Eqn. (3) the spin-torque comes out as $\vec{T} = -|\gamma|kM_s V\left[b(\hat{m}\times\hat{M}) + c(\hat{m}\times\hat{m}\times\hat{M})\right]$. which has the same form as the standard torque equations used extensively in literature [5, 6, 7].

Our formulation leads to a simple criterion for **coherent precession** which is considered one of the hallmarks of spin-torque. To see this we note that one can write Eqn. (1) in the following form

$$\frac{d\hat{m}}{dt} = -\frac{|\gamma|}{1+\alpha^2} \hat{m} \times (\hat{H}_{eff} + k(b-\alpha c)\hat{M})$$
$$-\frac{\alpha|\gamma|}{1+\alpha^2} \hat{m} \times \hat{m} \times (\hat{H}_{eff} + k(b+c/\alpha)\hat{M}) \quad (9)$$

Noting that coherent precession arises when the second term is zero, we obtain

$$\vec{H}_{eff} = -k\left(b + \frac{c}{\alpha}\right)\hat{M} \quad (10)$$

so that Eqn. (9) reduces to $\frac{d\hat{m}}{dt} = \frac{|\gamma|kc}{\alpha} \hat{m} \times \hat{M}$ yielding $|\gamma|kc/\alpha$ as the precession frequency. Since the "$c$" term is zero under equilibrium conditions (see **Appendix A**), coherent precession is possible only under non-equilibrium conditions, as one would expect.

**Nature of the pseudo field:**

Now that we have established the relationship between our concept of pseudo field and the standard form of torque having a Slonczewski and a field like term, let us try to examine the pseudo field more deeply. The first term in Eqn. (8) is in the same direction as the magnet. Hence this does not contribute anything to the torque and may be ignored. As for the second term, we see that if the coefficient $b$ were independent of $\theta$ and $\varphi$, $\vec{\nabla}_m \times (b\hat{M}) = 0$. This means that for the case when $b$ is independent of $\theta$ and $\varphi$, the second term acts as a conservative field. As for the third term in Eqn. (8) we show in **Appendix B**, that independent of the angular dependence of $c$, the third term always constitutes a non-conservative field. To summarize, the pseudo field that gives the spin transfer torque has two terms, one of which is in-plane with the magnets and may or may not be a conservative field. On the other hand, the second term is out-of-plane, is always non-conservative and can only appear at out-of-equilibrium.

## 4. Switching behavior in tunneling barrier based spin torque devices:

Let us now consider switching in tunneling barrier based spin torque devices. Our formulation is based on the coupled NEGF-LLG methodology described above. The details of NEGF implementation of transport for tunneling barrier based spin torque devices have been discussed in [16]. Here we shall skip the details and only present results. In brief, our formulation is based on effective mass description. We sum over the transverse modes assuming that the inter-mode coupling is negligible. Also, we only take the torque at the surface of the soft magnet. We have shown [16] that this methodology gives reasonable agreement with both the current and the tunneling magneto resistance (TMR) as a function of voltage by using effective mass and barrier height as fitting parameters. In this case, we shall use similar parameters as used in [16, 17]. A typical bias and angular dependence of $b$ and $c$ and the torque components are shown in Fig. 2. Note that the bias and angular dependence of the torque components show the same qualitative dependence as in the recent ab-initio study [10]. The bias dependence of $b$ and $c$ can be approximately written as $b = b_0 + b_1 V^2; c = c_1 V$.

Also, from the Fig. 2, it is evident that both $b$ and $c$ are completely independent of $\theta$ and $\varphi$ for the tunneling device as we have considered here. This means that the pseudo field will have a conservative part due to ($b\hat{M}$), where $b$ is symmetric with voltage. Fig. 3 shows the switching of magnetization with applied voltage. One would see that it takes less time to go from AP to P configuration compared to P to AP for the same magnitude of voltage. This means that it would take more voltage to switch from P to AP for a particular width of the voltage pulse. This result is surprising considering that both the torque magnitudes shown in Fig. 2 are completely symmetric with voltage. However the reason would be clear if we look at the pseudo field. As mentioned above, ($b\hat{M}$) is conservative and does not change polarity with voltage. This means that ($b\hat{M}$) acts as if an external magnetic field was applied in the direction of $\hat{M}$ irrespective of the voltage polarity. As a result, it directly changes the potential energy of the system helping the AP to P transition while acting against the P to AP switching.

An important thing to note is the fact that $b_0$, the equilibrium component of $b$, would also introduce an asymmetry in switching voltage and it manifests itself as an exchange field in the equilibrium R-H loops. However, the significance of $b$ being independent of angular position is that even if we compensate for this exchange field by making the equilibrium hysteresis loop completely symmetric, for example, by applying an external magnetic field, there will still be an asymmetry in the switching current due to $b_1$.

Notice that this asymmetry in the switching voltage is not dependent on the symmetric nature of $b$ shown in Fig. 2. As long as $b$ is not purely anti-symmetric, the effect remains. This asymmetry is also in addition to that arising from any voltage asymmetry in the magnitude of $c$, i.e., the in-plane torque component. It is worth mentioning, however, that two [6,7] of the three torque measurement experiments done so far have found $c$ to be anti-symmetric (making its magnitude symmetric) at least in the low voltage region in agreement with ab-initio calculation [10]. Our own calculations also support the anti-symmetric nature of $c$. This suggests that the dominant reason for the asymmetry in switching voltages for tunnel based devices [18] may arise from field like terms. This is surprising considering the fact that the field like term was minimal and was normally ignored in the earlier devices based on metallic channels.

## 5. Conclusion:

By formulating spin transfer torque as a pseudo field proportional to the spin resolved electron density, we have been able to show how the field like torque can introduce a voltage symmetric conservative torque on the magnet and thereby cause an asymmetry in the switching voltages for tunneling barrier based spin torque devices. It will be interesting to explore if this effect can be utilized to reduce the switching voltage by appropriate device design. Our results also suggest that that one should consider maximizing the electron density while exploring novel device designs [19, 20, 21] involving spin transfer torque. Furthermore, the ability to change the potential energy of a system (by virtue of a voltage induced conservative field [22]) may also have important implications for voltage induced energy conversion and phase transition.

## Appendix A: Proof of Eqn(s) (7), (8)

Let us assume that the fixed magnet $\hat{M}$ and the soft magnet $\hat{m}$ are both in the $\hat{x} - \hat{z}$ plane (see Fig. 1.) so that the Hamiltonian $H$ (see Eqn. (6)) completely real, assuming the vector potential $\vec{A}$ to be zero. It can then be shown that the Green's function is symmetric (Chapter 3, [15]): $G = G^T$. We shall use this symmetry property of the Green's function to understand the form of $\vec{\mu}$ which is defined in terms of correlation function $G^n$, $\vec{\mu}_\alpha \equiv -\int dV \int dE \, Trace\left[\vec{\sigma}_\alpha G^n\right]$ with $G^n$ given by [14]

$$G^n = A_1 f_1 + A_2 f_2 = \frac{1}{2}(A_1 + A_2)(f_1 + f_2) + \frac{1}{2}(A_1 - A_2)(f_1 - f_2) \qquad (A1)$$

where, $A_1$ and $A_2$ are the partial spectral functions due to contact 1 and 2 respectively $\left(A_{1,2} = G\Gamma_{1,2}G^\dagger\right)$. Now, both $A_{1,2}$ are Hermitian, but not symmetric, since $A_{1,2} = G\Gamma_{1,2}G^\dagger$, so that $A_{1,2}^T = \left(G^\dagger\right)^T \Gamma_{1,2}^T G^T = G^\dagger \Gamma_{1,2} G \neq A_{1,2}$. However, the total spectral function $A(= A_1 + A_2)$ can be written as $i(G - G^\dagger)$ and hence symmetric: $A^T = A$. This means that $A$ is purely real and can be expressed as

$$A = P\sigma_x + Q\sigma_z; \qquad (A2)$$

While $\qquad A_1 - A_2 = P'\sigma_x + Q'\sigma_y + R'\sigma_z;$

*At equilibrium,* only the $A_1 + A_2$ term in Eqn. (A1) is non-zero, so that the magnetization $\vec{\mu}$ can be written as stated in Eqn. (7) while *under non-equilibrium condition,* it has the more general form stated in Eqn. (8): $\vec{\mu} = a\hat{m} + b\hat{M} + c\left(\hat{m} \times \hat{M}\right)$.

## Appendix B: Non-Conservative Nature of Pseudo-field

In *Appendix A* we showed that the pseudo-field $\vec{\mu}$ lies entirely in-plane at equilibrium, but can have an out-of-plane component away from equilibrium. We will now show that at equilibrium it is conservative, but away from equilibrium, the out-of-plane component makes it non-conservative.

Assume that the fixed magnet $\hat{M}$ points along $\hat{z}$ (**Fig.1**) and the soft magnet points along $\hat{m}$ where $\hat{m} = [\sin\theta\cos\phi \quad \sin\theta\sin\phi \quad \cos\theta]$ defined in a spherical co-ordinate system. The other unit vectors can be written as $\hat{\theta} = [\cos\theta\cos\phi \quad \cos\theta\sin\phi \quad -\sin\theta]$ and $\hat{\phi} = [-\sin\phi \quad \cos\phi \quad 0]$. We can write the curl of the pseudo-field as

$$\vec{\nabla}_m \times \vec{\mu} = \frac{\hat{m}}{\sin\theta}\frac{\partial}{\partial\theta}\left(\sin\theta\mu_\phi\right) - \hat{\phi}\frac{\partial}{\partial\theta}\mu_m \tag{B.1}$$

where we have dropped terms involving $\partial/\partial m$, $\partial/\partial\theta$ since we assume $|m| = 1$, $\phi = 0$ to be fixed and only consider changes in the direction $\theta$ of the magnetization of the soft magnet relative to the fixed magnet ($\hat{z}$).

We write the pseudo-field as $\vec{\mu} = \mu[\sin\theta'\cos\phi' \quad \sin\theta'\sin\phi' \quad \cos\theta']$ so that we obtain (with $\phi = 0$),

$$\mu_m = \mu\left(\sin\theta\sin\theta'\cos\phi' + \cos\theta\cos\theta'\right) \tag{B.2a}$$

$$\mu_\theta = \mu\left(\cos\theta\sin\theta'\cos\phi' - \sin\theta\cos\theta'\right) \tag{B.2b}$$

$$\mu_\theta = \mu\sin\theta'\sin\phi' \tag{B.2c}$$

Now, if we change the $\theta$ of the soft magnet, its angle with the fixed magnet changes and in response the pseudo field could in general change arbitrarily making both terms in Eqn.(B.1) non-zero. But the component $\mu_m$ contributes nothing to the actual torque, and we could arbitrarily define it to be a constant so that the only non-zero curl arises from the first term:

$$\vec{\nabla}_m \times \vec{u} = \frac{\hat{m}}{\sin\theta} \frac{\partial}{\partial\theta} \left( \sin\theta \mu_\phi \right) \tag{B.4}$$

This means the curl is non-zero unless $\sin\theta\, \mu_\phi = \mu \sin\theta \sin\theta' \sin\phi'$ and it does not change as the $\theta$ of the soft magnet is rotated. This can happen only if $\mu_\phi$ is identically zero, which is exactly what happens under ***equilibrium*** conditions (see ***Appendix A***): the pseudo-field only has in-plane components, which means that $\phi' = \phi = 0$, making $\mu_\phi = 0$. Hence the pseudo-field is in-plane and conservative in equilibrium, but away from equilibrium it can have an out-of-plane component that will make it non-conservative.

**Figure Captions:**

**Fig. 1.** Schematic of tri-layer device. The left contact is the pinned ferromagnet having magnetization along the z-axis. The right contact is the free layer and the channel material is an oxide. $\hat{z}$ is the easy axis and $\hat{x} - \hat{z}$ is the easy plane. Transport occurs in y-direction. The device region is modeled using appropriate Hamiltonian, $H$, and electrostatic potential $U$ and the contacts are taken into account by self energy matrices $\Sigma_1$ and $\Sigma_2$, whose anti-Hermitian components $\Gamma_{1,2} = i(\Sigma_{1,2} - \Sigma_{1,2}^{\dagger})$ are broadening matrices due to contacts 1 and 2 respectively [14].

**Fig. 2. (a)** Typical variation of $b$ and $c$ as a function of voltage for tunnel based spin torque devices. $b$ shows symmetric and $c$ shows anti-symmetric voltage dependence. **(b)** Bias dependence of in-plane and out-of-plane components of Torque for tunnel based spin torque devices. **(c), (d)** The variation of $b$ and $c$ as a function of $\theta$ at a fixed $\phi$ and as a function of $\phi$ at a fixed $\theta$ respectively at a fixed voltage for a tunnel based spin torque device. We see that $b$ and $c$ at a fixed voltage are independent of both $\theta$ and $\phi$. **(e)** Typical variation of differential torque (w.r.t. voltage) as a function of the relative angle $\theta$ between the magnetizations of the ferromagnetic electrodes.

**Fig. 3.** The switching dynamics with same voltages with opposite polarity: positive voltage for AP to P and negative voltage for P to AP. For clarity, we have only marked the z component with bold blue color. The dashed curve shows AP to P and the solid curve shows P to AP transitions. **(a)** For the same voltage amplitude, the AP to P transition is faster than P to AP. Note the dashed line where the AP to P transition is almost complete while the P-to-AP transition is just around its half-way mark. **(b)** To get a symmetric switching time, it takes almost 30% more voltage (V-) for P-to AP compared to the AP-to-P transition. No external magnetic field has been assumed.

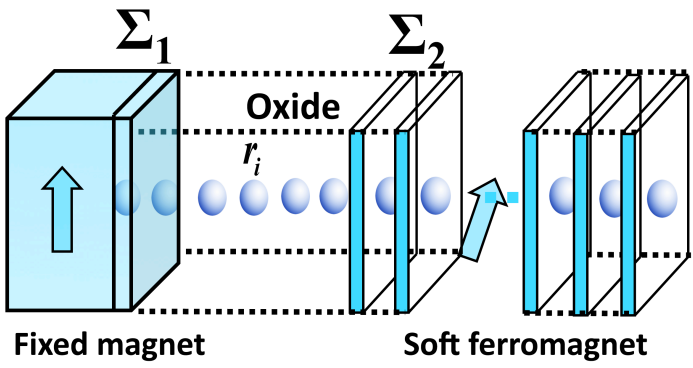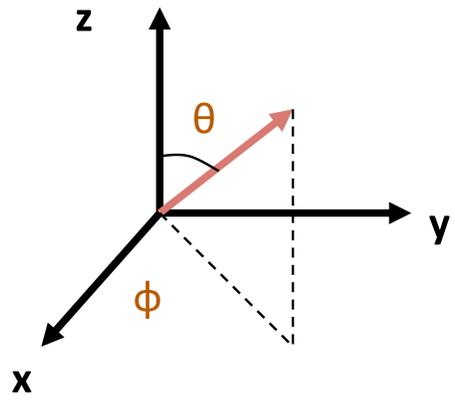

**Fig. 1**

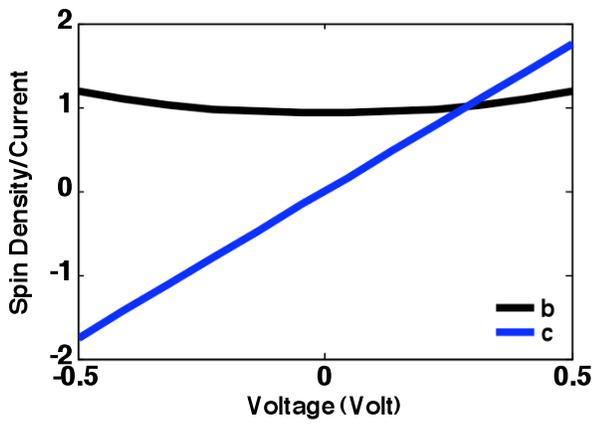

**Fig. 2 (a)**

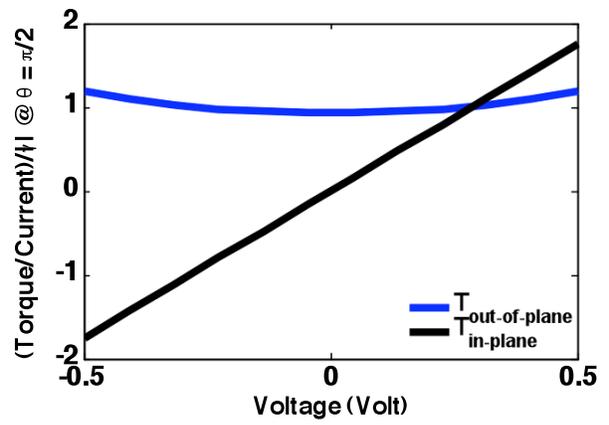

**Fig. 2 (b)**

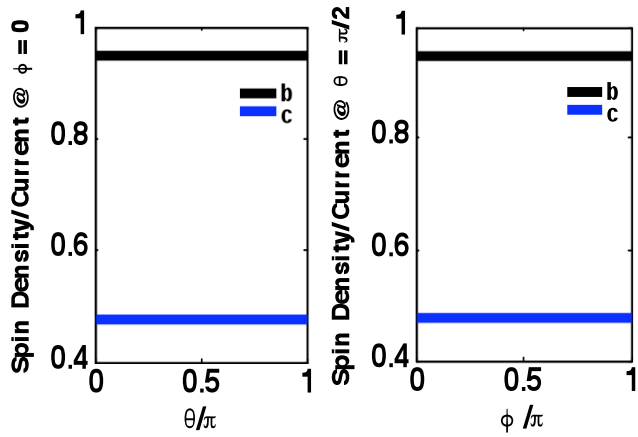

**Fig. 2 (c)**       **Fig. 2 (d)**

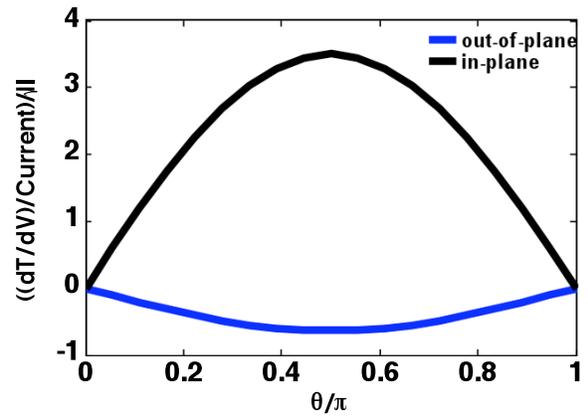

**Fig. 2 (e)**

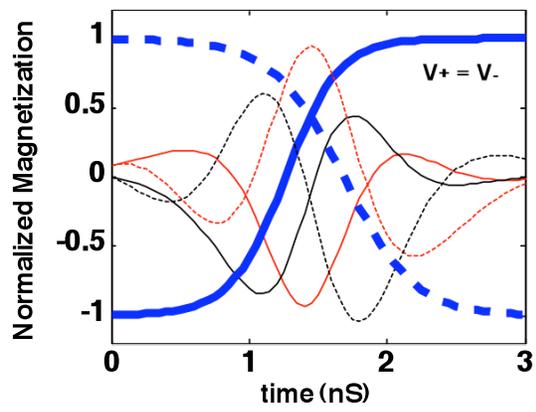 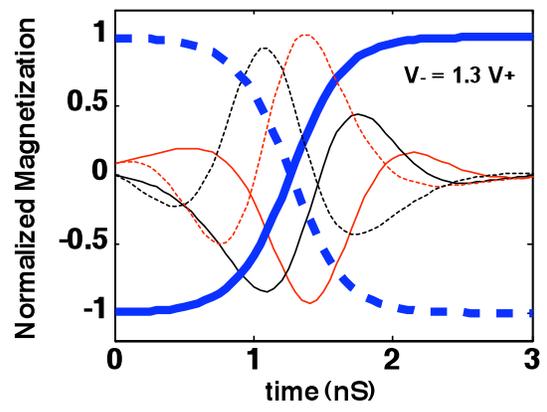

**Fig. 3 (a)**     **Fig. 3 (b)**